# Astro2020 Science White Paper

# Searches for Technosignatures in Astronomy and Astrophysics

**Primary thematic area: Planetary Systems**, especially exobiology and the search for life beyond the Solar System. **Secondary thematic areas**: •Star and planet formation •Resolved stellar populations and their environments • Galaxy evolution

**Principal Author:  Jason T. Wright**
**Email:** astrowright@gmail.com **Phone:**(814) 863-8470
525 Davey Laboratory
Department of Astronomy & Astrophysics & Center for Habitable Worlds
Penn State University, University Park, PA 16802

**42 Endorsers / co-signers:**
**Lori Allen**, National Optical Astronomy Observatory; **Daniel Angerhausen**, Center for Space and Habitability, Bern U., Switzerland & Blue Marble Space Institute of Sciences, Seattle, USA; **David Ardila**, Jet Propulsion Laboratory, Caltech; **Amedeo Balbi**, Università degli Studi di Roma "Tor Vergata"; **Anamaria Berea**, U. of Central Florida; **Alan Boss**, Carnegie Institution; **Tabetha Boyajian**, Louisiana State U.; **Douglas A. Caldwell**, SETI Institute; **P. Wilson Cauley**, LASP, U. of Colorado Boulder; **William Cochran**, U. of Texas at Austin; **Steve Croft**, UC Berkeley; **James R. A. Davenport**, U. of Washington; **Julia DeMarines**, UC Berkeley; **Kathryn Denning**, York U.; **Chuanfei Dong**, Princeton U.; **J. Emiio Enriquez**, UC Berkeley; **Theresa Fisher**, Arizona State U. **Adam Frank**, U. of Rochester; **Dawn M. Gelino**, NASA Exoplanet Science Institute, IPAC, Caltech; **Claudio Grimaldi**, Ecole Polytechnique Fédérale de Lausanne; **David H. Grinspoon**, Planetary Science Institute; **Nader Haghighipour**, Institute for Astronomy, U. of Hawaii; **Hilairy Hartnett**, Arizona State U.; **Jacob Haqq-Misra**, Blue Marble Space Institute of Science; **Paul Kalas**, UC Berkeley; **Shubham Kanodia**, Pennsylvania State U.; **Eric J Korpela**, UC Berkeley; **Joseph Lazio**, Jet Propulsion Laboratory, Caltech; **Jacob Lustig-Yaeger**, U. of Washington, Seattle; **Jean-Luc Margot**, UCLA; **Michael W. McElwain**, NASA Goddard Space Flight Center; **Henry Ngo**, National Research Council Canada; **Sarah Rugheimer**, U. of Oxford; **Caleb Scharf**, Columbia U.; **Edward W. Schwieterman**, UC Riverside; **Hector Socas-Navarro**, Instituto de Astrofísica de Canarias; **Angelle Tanner**, Mississippi State U.; **Jill Tarter**, SETI Institute; **Gabriel G. De la Torre**, Universidad de Cádiz; **Sara Walker**; Arizona State U. **Shelley A. Wright**, UC San Diego; **Abel Méndez**, PHL @ UPR Arecibo; **Frank Soboczenski**, King's College London

# New Opportunities for Searches for Technosignatures

The search for life beyond the Solar System—a major part of the Planetary Systems thematic area of the Astro2020 Decadal process—includes the search for technological life. Although financial support for such searches at the NSF and NASA and in past decadal surveys has been weak to nonexistent, recent advances in astrobiology, astrophysics, and advances in technical capability make searches for technosignatures[1] a compelling theme for 2020-2030 and beyond.

The discovery of the ubiquity of exoplanets from 1995-2010 led to a significant increase in the expectation value for the number of sites of potential life in the Galaxy. At the beginning of that decade, we understood the number of Earth-like planets per star, to be $\sim 0.25$ (e.g., Howard et al. 2010). The launch of *Kepler* in 2009 led to a second revolution in exoplanetary astronomy, including the discovery of thousands of new rocky planets including many receiving stellar flux at levels consistent with the presence of liquid surface water. We now know that potential sites of abiogenesis in the Milky Way abound. **With this understanding, the search for life beyond the solar system has rightly intensified.** Searches for sites of potential life and planets amenable to atmospheric characterization has driven much of exoplanetary astronomy, and missions such as *JWST* explicitly include the search for life among their science goals.

So it is not surprising that simultaneously, the past 10 years have seen a resurgence in searches for technosignatures (and philanthropic efforts such as Breakthrough Listen Initiative have played no small part in this change). Some of the drivers for this include:
- New developments in radio and optical/NIR instrumentation,
- The advent of the *WISE* all-sky catalogs
- The repurposing of ground-based data collected for other reasons, e.g., radial velocity planet search data used for laser searches
- The promise of the new 30 m class observatories
- The stellar photometric revolution ushered in by *Kepler*
- Opportunities for big data analysis via new machine learning and other algorithms
- The promise of LSST and other extremely high data rate projects
- The dawn of multi-messenger astronomy

Many of the arguments for inclusion of searches for technosignatures among the scientific priorities of NASA and the NSF were presented at the NASA Technosignatures Workshop held in Houston in 2018 in anticipation and preparation for passage of Congressional authorization language, which would have required NASA to report on its technosignature search plans.[2] Note that contrary to popular understanding, *there is no statutory prohibition against NASA funding SETI work; the reasons for lack of funding are purely cultural* (see, e.g., Garber 1999).

---

[1] "Technosignatures", by analogy with biosignatures, are signs of technological life, including communicative transmissions and other technologies. In general, "the search for technosignatures" is used synonymously with "SETI" (Tarter 2007, Wright et al. 2018).

The NASA Technosignatures Workshop produced a report to NASA (Technosignature Workshop Participants, 2018) addressing five major question areas defined by NASA:

1. *What limits do we currently have on technosignatures?*
2. *What is the state-of-the-art for technosignature detection?*
3. *What assets are in place that can be applied to the search for technosignatures? What planned and funded projects will advance the state-of-the-art in future years, and what is the nature of that advancement?*
4. *What new surveys, instruments, technology development, data-mining algorithms, theory and modeling, etc., would be important for future advances in the field?*
5. *What role can NASA partnerships with the private sector and philanthropic organizations play in advancing our understanding of the technosignatures field?*

While the report makes no recommendations, it does identify several areas in which NASA can participate and support the search for technosignatures. **We endorse this report and recommend that the opportunities it describes be identified by the Decadal survey as components of a compelling scientific theme for 2020–2030 and beyond for both NASA and the NSF.** Many of the specific critical questions and opportunities, as well as the observational and theoretical advances that will address them, are described in other white papers, submitted to the Decadal call, described below.

## Technosignatures and Biosignatures

**The search for technosignatures logically shares a continuum with the search for biosignatures, and is properly considered part of NASA and the NSF's science portfolios. Here we endorse one such articulation of this principle, the Wright (2018) white paper.**

---

[2]H.R. 5530 §311(c). Both this and the Senate 2018 NASA authorization bill included language specifying that technosignatures are part of NASA's mission to find life in the universe. Neither bill was passed by the 115th Congress, however the language had significant stakeholder buy-in and such bills serve as markers for the next Congress's authorization work. Both the Houston Workshop report and the Astro2020 Decadal document will serve as roadmaps for NASA should such language become law. Separately in 2019, *appropriations* language appropriating $10M/yr for NASA to "partner with private sector and philanthropic organizations" on searches for technosignatures did not survive committee.

Briefly, the argument for searches for technosignatures is at least as strong as that for searches for biosignatures. Technosignatures may be more ubiquitous than biosignatures (since technological life can, in principle, spread among the stars on its own volition), may be more obvious than biosignatures (since technology can, in principle, extend to planetary, stellar, or even galactic scales and exploit far larger energy supplies than a biosphere can), and may be more unambiguous (for instance if a communicative signal were detected).

Many of the objections to the search for technosignatures (e.g., claims that the search is ill-defined) are shared by the searches for biosignatures, and are reasons to refocus research energies on those topics, not to avoid them. As with the search for biosignatures, searches for technosignatures begin with explorations of where particular sorts of life might exist, and continue with explorations of what signs that life might exhibit, how we might detect those signs, how we might be confused by abiotic phenomena, and finally a search for those signs.

A common misconception is that searches for technosignatures are a distinct activity from the rest of astronomy or astrobiology. But just as searches for biosignatures require an interdisciplinary approach that both is informed by and advances, for instance, biology, planetary science, exoplanetary science, and stellar astrophysics, searches for technosignatures require detailed understanding of confounding effects of exoplanets, stars, and galaxies, and false positives often prove to be extremely interesting natural objects in their own right.

There is also an erroneous perception that searches for technosignatures have been thoroughly pursued, and that technosignatures have been shown not to exist in great numbers. On the contrary, Wright, Kanodia, & Lubar (2018) have shown, consistent with previous work by Tarter (2010) and Gray et al. (2002), that despite the high public profile of the field very little actual searching has occurred, even in the better- (but still under-)funded field of narrowband radio SETI work.

## Technosignature Desiderata

Searches for technosignatures include expansions of existing radio and optical SETI work and also many more projects. When determining which technosignatures are the best targets for searches, one must optimize among many axes of merit. Some such axes are illustrated in Figure 1, presented at the NASA Technosignatures Workshop.

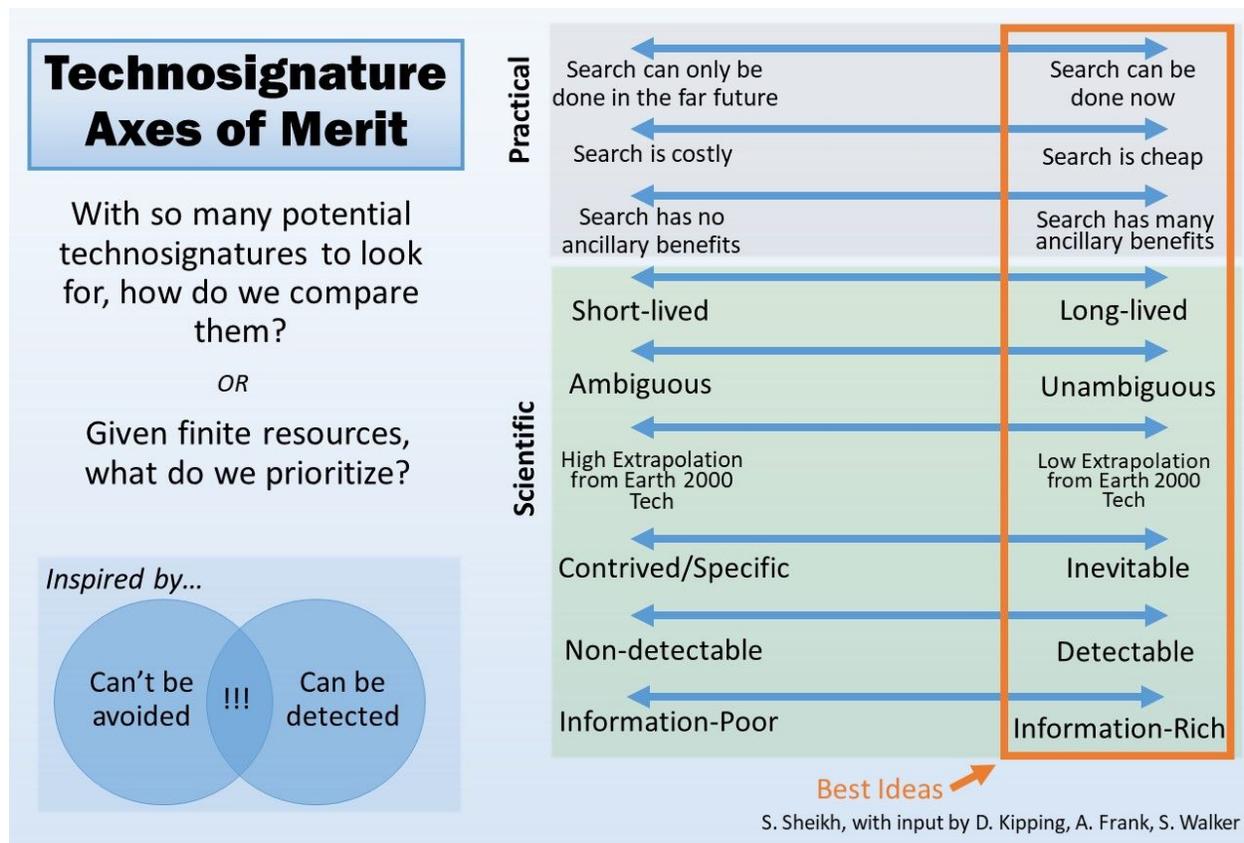

Figure 1: Technosignature Axes of Merit, illustrating some of the considerations that go into developing a good search strategy for technosignatures. From the Houston Report.

These axes are:
1. How much technology development is required by us to perform the search.
2. How expensive, in dollars, time, and opportunity cost, the search is.
3. How many ancillary benefits a search generates, e.g. from development of methods and technologies applicable elsewhere or discovery of novel astrophysics or types of objects.
4. How long-lived the purported technosignature is or, equivalently, the duty cycle of the signature over a civilization's lifetime.
5. How unambiguously the technosignature can be distinguished from natural confounders.
6. How much technology development is required by the technological civilization beyond "Earth 2000" levels to produce the technosignature.
7. How contrived or specific the motivations and capabilities of a technological civilization must be to produce the technosignature.
8. How easily detected the technosignature is.
9. How much information a technosignature contains. Radio signals might be information-rich; thermal emission might contain nearly zero information. Very distant or faint sources might offer little or no opportunity to learn more about the technology.

## Contributions to Broader Understanding and Specific Opportunities

The discovery of life elsewhere in the universe is a primary goal of NASA, and the implications of finding *technological* life are even greater than finding purely microbial life.

But even before they are successful, such searches have implications across astronomy and beyond. Exactly because SETI seeks signals of obviously artificial origin, it must deal with and examine the rare and poorly understood astrophysical phenomena that dominate its false positives (e.g., pulsars, AGN, and even Boyajian's Star, Boyjajian et al. 2016). Anomalies discovered during searches for pulsed and continuous laser emission (Howard et al. 2007, Wright et al. 2014, Tellis & Marcy 2015, 2017), radio signals, large artificial structures (Dyson 1960, Griffith et al. 2015, Wright et al. 2016), and other astrophysical exotica push astrophysics in new and unexpected directions. The new technology and search strategies developed for SETI can also be applied to other fields (e.g., Gajjar et al. 2018).

Indeed, the NASA Technosignatures Workshop emphasized the wide variety of ways NASA can support the search for technosignatures. SETI is an interdisciplinary field (Cabrol 2016) and even beyond the potential for NASA's Deep Space Network to play an important role in the radio component of SETI, archival data from NASA assets have played an important role in SETI for decades: from Solar System SETI using interplanetary cameras, to waste heat searches using *IRAS* (Carrigan 2009), *WISE*, *Spitzer*, and *GALEX* (Griffith et al. 2015), to searches for artifacts with *Kepler* (Wright et al. 2016) and *Swift* (Meng et al. 2017). Future ground-based projects like *LSST* and space-borne projects like *JWST* and *WFIRST* will undoubtedly provide additional opportunities for SETI research, both as ancillary output of legacy and archival programs and through independent SETI projects in their own right.

Some of the technosignatures searches ripe for investment and development include:
1. Radio technosignatures (See white paper by **J.-L. Margot**)
2. Continuous and pulsed laser emission
3. Thermal infrared technosignatures (See white paper by **J. Wright**)
4. Megastructures in transit (See white paper by **J. Wright**)
5. Applications of data science (See white paper by **A. Berea**)
6. Scientific implications of (non-)detection (See white paper by **J. Haqq-Misra**)
7. Observing the Earth as a communicating exoplanet (See white paper by **J. DeMarines**)
8. Multi-messenger SETI
9. SETI in the Solar System